\documentclass[aps,prb,twocolumn]{revtex4}
\usepackage{graphicx}
\usepackage{dcolumn}
\usepackage{amsmath}
\usepackage{amssymb}
\usepackage{color}
\usepackage{subfigure,amsmath,verbatim,moreverb,bm}
\def\bef{\begin{framed}}
\def\eef{\end{framed}\noindent}
\def\be{\begin{equation}}
\def\ee{\end{equation}}
\def\ber{\begin{eqnarray}}
\def\eer{\end{eqnarray}}
\def\xiv{{\boldmath{\xi}}}

\def\fbold{\mbox{\boldmath ${\cal F}$}}
\def\nablabold{\mbox{\boldmath $\nabla$}}
\def\rv{{\bf r}}
\def\jv{{\bf j}}

\def\Av{{\bf A}}
\def\Bv{{\bf B}}

\def\xiv{{\mbox{\boldmath $\xi$}}}

\def\uv{{\bf u}}
\def\vv{{\bf v}}
\def\xv{\rv}

\def\calBv{\mbox{\boldmath ${\cal B}$}}
\def\calKv{\mbox{\boldmath ${\cal K}$}}
\def\calOv{\mbox{\boldmath ${\cal O}$}}

\def\nn{\nonumber}

\def\fbold{\mbox{\boldmath ${\cal F}$}}
\def\Mbold{\mbox{\boldmath ${\cal M}$}}

\begin{document}
\title{Quantum continuum mechanics in a strong magnetic field}
\author{S. Pittalis}
\author{G. Vignale}
\affiliation{Department of Physics, University of Missouri-Columbia,
Columbia, Missouri 65211}
\author{I. V. Tokatly}
\affiliation{ETSF Scientific Development Centre, 
Departamento de F\'isica de Materiales, Universidad del Pa\'is Vasco UPV/EHU, Av. Tolosa 72, E-20018 San Sebasti\'an, Spain}
\affiliation{IKERBASQUE, Basque Foundation for Science, E-48011, Bilbao, Spain}
\date{\today}
\begin{abstract}
We extend a recent formulation of quantum continuum mechanics [J. Tao et. al, Phys. Rev. Lett. {\bf 103}, 086401 (2009)]  to many-body systems subjected to a magnetic field.  To accomplish this, we propose a modified Lagrangian approach, in which motion of infinitesimal volume elements of the system is referred to the ``quantum convective motion" that the magnetic field produces already in the ground-state of the system. In the linear approximation, this approach results in a redefinition of the elastic displacement field  $\uv$, such that the  particle current $\jv$ contains both an electric displacement and a magnetization contribution: $\jv=\jv_0+n_0\partial_t \uv+\nabla \times (\jv_0\times \uv)$,  where $n_0$ and $\jv_0$ are the particle density and the current density of the ground-state and $\partial_t$ is the partial derivative with respect to time.  
In terms of this displacement,  we formulate an ``elastic approximation" analogous to the one proposed in the absence of magnetic field.  
The resulting equation of motion for $\uv$ is  expressed in terms of ground-state properties -- the one-particle density matrix and the two-particle pair correlation function -- and in this form it neatly generalizes the equation obtained for vanishing magnetic field.
\end{abstract}
\pacs{ ... }
\maketitle

\section{Introduction}\label{I}

Consideration of an hydrodynamical formulation of the electron dynamics goes back to the early days of quantum mechanics~\cite{Madelung, Bloch}. In the modern language of
time-dependent (current-)density-functional theory~\cite{TDDFT} (TD(C)DFT), this naturally leads to appealing time-dependent orbital-free methods~\cite{GhoshDeb, Zaremba}. 
In this spirit, in two recent papers~\cite{PRLCM,CM}, the problem of calculating the linear response of a generic quantum many-body systems to an external  time-dependent potential  has been reformulated in the language of  quantum continuum mechanics (QCM).   
In this approach, the non-equilibrium state of the system is described in terms of an elastic ``displacement field"  $\uv(\rv,t)$  (a function of cartesian coordinates $\rv$ and  time $t$),   such that an infinitesimal volume element of the system that is  located at point $\rv$ in the equilibrium state will be located at $\rv+\uv(\rv,t)$ in the non-equilibrium state.  
The particle current density is connected to the displacement by the relation
\be\label{Displacement1}
\jv(\rv,t)=n_0(\rv)\partial_t \uv(\rv,t)\,,
\ee
where $\partial_t$ represents a partial derivative with respect to time and $n_0(\rv)$ is the ground-state density.   An ``elastic approximation" was then introduced, based on the idea that the time evolution of the wave function can be described as a geometric deformation of the ground-state wave function, the deformation being defined by the displacement field $\uv$.  
More precisely, the wave function of the deformed state $|\psi[\uv]\rangle$ was expressed in terms of the ground-state wave function $|\psi_0\rangle$ in the following manner:
\be\label{DeformedPsi}
|\psi[\uv]\rangle = \exp\left[-i\int d\rv ~ \hat \jv(\rv)\cdot\uv(\rv)\right]|\psi_0\rangle\,,
\ee
where $\hat \jv(\rv)$ is the canonical current density operator acting as the generator of differential translations -- a different translation at each point in space.
(Here and in the following, we set the mass of the particles $m = 1$).  Starting from Eq.~\eqref{DeformedPsi}, a closed equation of motion for $\uv$ could be derived, which is demonstrably exact for (i) one-particle systems, and (ii) many-particle systems subjected to high-frequency fields.  This equation has the form
\ber\label{eom.simple} 
 n_0(\rv)\partial_t^2 \uv (\rv,t) &=&- n_0(\rv) \uv \cdot\nablabold (\nablabold V_0) -\frac{\delta T_2[\uv]}{\delta \uv (\rv,t)}-\frac{\delta W_2[\uv]}{\delta \uv (\rv,t)}\nn\\
&&-n_0(\rv)\nablabold V_1(\rv,t)\,,
\eer
where $V_0(\rv)$ is a static potential that defines the nature of the many-body system, $V_1(\rv,t)$ is the small time-dependent potential to which the many-body system responds,  $T_2[\uv]$ is  the kinetic energy density of the deformed state \eqref{DeformedPsi} expanded to second-order in $\uv$, and $W_2[\uv]$ is the electron-electron interaction energy density of the deformed state, also expanded  to second-order in $\uv$.  The functional derivatives that appear on the right hand side of Eq.~(\ref{eom.simple}) are actually linear integro-differential operators acting on $\uv$.  The explicit form of these operators was derived in Refs.~\onlinecite{PRLCM,CM}.  The final expressions involve only the following ground-state properties: the one-particle density matrix and the two-particle correlation function,
which are quantities that may be computed by means of quantum Monte Carlo methods~\cite{QMC}.

Quantum continuum mechanics holds great promise as a tool for simplifying and streamlining the calculation of excitation spectra, particularly in situations in which the ground-state correlations are well understood and the spectrum is dominated by collective excitations.
Another interesting possibility is to use quantum continuum mechanics as a tool for efficiently  approximating the density-density response function of the non-interacting Kohn-Sham system.  This possibility has been recently pursued by Gould and Dobson.\cite{Gould11}  The Kohn-Sham response function is then used by these authors, in combination with the random phase approximation, to generate more accurate exchange-correlation energy functionals, which capture dispersion forces between metals and insulators.
Moreover, the use of quantum continuum mechanics has allowed to better understand and simplify the derivation of the expression of
the high-frequency limit of the {\em exact} exchange-correlation kernel of TD(C)DFT~\cite{NazarovHFL}.
We should also mention that, another interesting approach that holds promise to overcome present limitations of  the available {\em approximate} KS  methods is the semiclassical density-matrix time dependent propagation
of Refs.~\cite{Maitra,Elliot}.

The theory of Refs.~\onlinecite{PRLCM,CM} was based on the assumption that the ground-state of the many-body system is time-reversal invariant and has no spin-orbit coupling, so that the ground-state current-density $\jv_0(\rv)=0$.
This paper is concerned with the extension of that theory to many-electron systems in the presence of a static magnetic field $\Bv_0(\rv) = \nablabold \times \Av_0(\rv)$, where $\Av_0(\rv)$ is a static vector potential.  
Thus, we  consider a system of $N$ identical particles described by the time-dependent Hamiltonian
\begin{eqnarray} 
\hat H(t) &=& \sum_{j = 1}^{N}\left[\frac{( -i \nabla_{j} 
+ \Av_0(\rv_{j}))^{2}}{2} + V_0(\rv_{j})+V_1(\rv_{j},t)\right] \nn \\ 
&+& \frac{1}{2}\sum_{j\ne k} W(|\rv_{j}-\rv_{k}|)
\label{H}
\end{eqnarray}
where $W(|\rv-\rv'|)$ is the electron-electron interaction potential (we set $e= \hbar=c=1$).
The time-dependent many-body wave function $\Psi(\rv_{1},\dots,\rv_{N},t)$
is the solution of the Schr\"odinger equation 
\begin{equation}
 i\partial_t\Psi(\rv_{1},\dots,\rv_{N},t) = H\Psi(\rv_{1},\dots,\rv_{N},t)
\label{SE}
\end{equation}
with initial condition 
\begin{equation}
\Psi(\rv_{1},\dots,\rv_{N},0) = \Psi_0(\rv_{1},\dots,\rv_{N})\,.
 \label{InitialPsi}
\end{equation}

There are  compelling reasons for working out the generalization of QCM to systems subjected to magnetic fields.  As discussed in Refs. \onlinecite{PRLCM,CM}, the elastic approximation is, in essence,  a collective approximation for inhomogeneous systems.  It condenses the  excitation spectrum of the many-electron system into a simpler spectrum of collective excitations, which still carry the exact full spectral strength (this is a consequence of the exactness of the theory in the high-frequency limit).  Naturally, such an approximation becomes more trustworthy when the excitations under study are  truly collective, as opposed to incoherent single-particle excitations.  But, it is well known that a strong magnetic field, by quenching the kinetic energy of an electronic system,   suppresses single-particle behavior and promotes collective behavior.  Indeed, one of the first successful theories of the excitation spectrum of the homogeneous two-dimensional electron gas (2DEG) at high magnetic field was based on a single-mode approximation very similar to our elastic approximation 
\cite{GMP}.  Nanopatterned electronic systems at high magnetic field (e.g., quantum dots) also exhibit strongly collective behavior.  
One can, for example, fabricate a lattice of quantum dots (nanopillars) by chemical etching on a 2DEG  \cite{AG1}
and observe the collective excitations of the resulting electronic system by Raman scattering \cite{AG2}.  By doing this experiment, new collective modes have been recently discovered \cite{AG2}, 
which have no counterpart in the homogeneous 2DEG.  We believe that our continuum mechanics will be useful precisely for a microscopic study of these collective modes. 

The generalization of Eq.~(\ref{eom.simple}) to systems described by the Hamiltonian~(\ref{H})  is not as straightforward as one might  initially think.  Of course, the presence of the vector potential modifies the form of the kinetic energy and introduces a Lorentz force term, but this is not the main difficulty.
The main difficulty arises from the fact that the ground-state current no longer vanishes: it has a finite expectation value,  $\jv_0(\rv)$.    This introduces an ambiguity in the definition of the displacement field.  

The first  possibility we explored is to define $\uv$ in close analogy to Eq.~(\ref{Displacement1}) such that
\be\label{Displacement1.1}
\jv(\rv,t)=\jv_0(\rv)+n_0(\rv)\partial_t\uv(\rv,t)\,.
\ee
With this definition, it is possible to derive within standard linear response theory, a closed equation of motion for $\uv$.  This will be done in Section II.   While this equation is formally elegant and gives some insight into the general properties of the solutions, it is very difficult to put it in an explicit and therefore useful form.  The reason is that one needs to calculate the ground-state expectation value of complicated commutators: the amount of algebra involved is formidable.  By contrast, in the treatment of the zero field case we could rely on a direct calculation of the energy of the deformed state \eqref{DeformedPsi} -- a comparatively simpler task that did not require the evaluation of complicated commutators.

Eq.~(\ref{Displacement1.1}) assumes that the excess current $\jv-\jv_0$ is entirely due to the time derivative of the displacement field, but, in the presence of a magnetic field, even a time-independent displacement can produce an excess current (see below).  After all, the ground-state is perfectly stationary, and yet it does carry a current $\jv_0$.  However, at variance with a time-dependent current, the current associated with a static deformation must necessarily have vanishing divergence in order to satisfy the continuity equation.  Thus, for example,  one must have $\nablabold \cdot \jv_0=0$.  Taking into account this condition, we are free to  add to the right hand side of  Eq.~(\ref{Displacement1.1}) the curl of a ``magnetization field".
In other words, we expect the most general form of the relation between current and displacement to have the form
\be\label{Displacement2}
\jv(\rv,t)=\jv_0(\rv)+n_0(\rv)\partial_t\uv(\rv,t)+\nablabold \times {\Mbold}(\rv,t)\,,
\ee
where ${\Mbold}(\rv,t)$ is a functional of $\uv$.   The divergence of the last term on the right hand side is guaranteed to be zero.  This is completely analogous to the material current in electrodynamics, which is customarily written as the sum of the time derivative of the electric polarization and the curl of the magnetization.~\cite{JDJ}

To determine the form of ${\Mbold}(\rv,t)$ up to the linear order in $\uv$, we assume that (i) ${\Mbold}(\rv,t)$ is a local functional of the displacement, i.e., it depends on $\uv(\rv,t)$ at the same point of space and time, (ii) A uniform displacement $\uv(\rv)=\uv$ must cause the ground-state current to be rigidly displaced by $\jv_0(\rv) \rightarrow \jv_0(\rv -\uv)$, i.e. we must have  $\nablabold \times {\Mbold}(\rv,t) = - (\uv \cdot \nablabold)\jv_0(\rv)$ for uniform $\uv$.  This  condition implies that ${\Mbold}(\rv) =  \uv \times \jv_0(\rv)$ and fixes the relation between current and displacement in the form
\be\label{Displacement2.2}
\jv(\rv,t)=\jv_0(\rv)+n_0(\rv)\partial_t\uv(\rv,t)+\nablabold \times [\uv(\rv,t) \times  \jv_0(\rv)]\,.
\ee

Remarkably, this relation emerges naturally from the Lagrangian formulation of the problem, which we present in Section III.    We remind the reader that in Refs.~\onlinecite{PRLCM,CM} the elastic approximation was derived via a transformation to a {\em local} non-inertial reference frame -- the so-called co-moving frame -- in which the density is constant and the current density is zero.  To achieve these conditions, the displacement field $\uv$, which defines the transformation to the co-moving frame, was related to the current via Eq.~(\ref{Displacement1}).  In the present situation, the physically ``natural" requirement for the co-moving frame is that the density  remain constant and the current density remain equal to the ground-state value $\jv_0$.  In other words, an observer ``riding" on a volume element should not detect any change in the density or the current density.  We found that this requirement determines the relation between the displacement field and the current density in the form of Eq.~(\ref{Displacement2.2}) just as we found from the heuristic argument given above.

Adopting the mentioned special co-moving frame is the crucial insight that  allows us to arrive at an explicit equation of motion for $\uv$ in the presence of a magnetic field.
Throughout the paper, we will consider a general situation of {\em non-collinear} and {\em non-uniform} magnetic fields (neglecting spin-degrees of freedom).
The equation of motion in the practically most common case of a {\it uniform} magnetic field is obtained from the equation of  motion ~(\ref{eom.simple}) by the following simple replacements:
\begin{enumerate}
\item Replace the time derivative $\partial_t$ by the ``convective derivative"
\be
D_t = \partial_t+\vv_0\cdot\nablabold
\ee
where $\vv_0=\jv_0/n_0$.
\item Include on the right hand side a ``Lorentz force term"
\be
 D_t \uv  \times  \Bv_0  \,.
\ee
\item Calculate $T_2[\uv]$ and $W_2[\uv]$ respectively from the expectation values of the kinetic energy and electron-electron interaction energy operators evaluated in the deformed ground-state wave function defined, just as in Eq.~(\ref{DeformedPsi}).
The electron-electron interaction energy term remains formally unaffected, while the kinetic energy term is modified by the inclusion of an effective vector potential  
accounting for the external  static vector potential and the corresponding contribution of the convective motion in the ground-state (remarkably, this modification 
vanishes for one-electron systems).
\end{enumerate}

The paper is organized as follows: 

In Section~\ref{II} we derive an approximate equation of motion for the conventional displacement field from the high-frequency expansion of the exact  current-current response function.   We show that this approximation is exact for one-particle systems and discuss the difficulties arising when one attempts to explicitly evaluate the formal expressions appearing in this equation.  

In Section~\ref{III} we address the difficulties discussed in  Section~\ref{II} by resorting to a  non-standard Lagrangian formulation  appropriate for systems in magnetic field.   To this end, we formulate the quantum many-body dynamics in a special co-moving frame, such that both the density and the current density retain their initial (ground-state) values at all times.  We arrive at an exact, but still not explicit equation of motion for the displacement field in terms of the Hamiltonian in the stress tensor of the co-moving frame.

In Section~\ref{IV} we introduce the  elastic approximation and obtain a closed, fully nonlinear equation of motion for the displacement field in this approximation.

Finally, in Section~\ref{V}, we linearize the elastic equation of motion and obtain an explicit, linear equation of motion for the displacement field.  It is shown that the displacement field obtained in this manner is not the conventional one, but  is related to the current density by Eq.~(\ref{Displacement2}).  

Section ~\ref{VI} concludes the paper with a  summary of the main results.

\section{Derivation from linear response theory}\label{II}
A formally exact equation of motion for the current density response of a many-particle system in the linear response regime can be easily derived from standard linear response theory.~\cite{gvbook,keldysh}
To this end, it is convenient to replace the external potential $V_1(\rv,t)$ by a time-dependent vector potential $\Av_1(\rv,t)=-\int_{-\infty}^t \nabla V(\rv,t') dt'$, which is physically equivalent since it gives rise to the same time-dependent electric field and no time-dependent magnetic field.  Assuming further that the time-dependence of the external field is periodic with angular frequency $\omega$ and switched on adiabatically at $t=-\infty$  with the system initially in the ground state $|\psi_0\rangle$, we obtain the standard result for the Fourier component of the current density at frequency $\omega$: 
\be\label{LinearResponse1}
j_{1,\mu}(\rv,\omega) =  \int d\rv' \chi_{\mu\nu}(\rv,\rv',\omega)\frac{\partial_\nu V_{1}(\rv')}{i \omega}\,,
\ee
where $\chi_{\mu\nu}(\rv,\rv',\omega)$ is the current-current response function (Einstein summation convention is used throughout the paper) 
\ber\label{CHI}
\chi_{\mu\nu}(\rv,\rv',\omega) &=& n_0(\rv)\delta(\rv-\rv')\nn\\
&&-i\int_0^\infty dt  e^{i\omega t}\langle \psi_0|[\hat \jv(\rv,t),\hat\jv(\rv',0)]| \psi_0\rangle\nn\\
\eer
and
$\frac{\partial_\nu V_{1}(\rv')}{i \omega}$ is the Fourier component of $\Av_1(\rv,t)$ at frequency $\omega$.   
In the above equation~(\ref{CHI}) $\hat\jv(\rv,t)$ is the current-density operator defined as
\be\label{jp}
{\hat \jv}(\rv) = \frac{1}{2}\sum_{j=1}^N \left\{[-i \hat{\nabla}_j + \Av_0(\hat{\rv}_j)],\delta(\rv - \hat{\rv}_j)\right\}\,,
\ee
where $\{\hat{A},\hat{B}\}=\hat{A}\hat{B}+\hat{B}\hat{A}$ is the anticommutator.  This operator evolves in time under the unperturbed Hamiltonian, i.e., the Hamiltonian (\ref{H}) without the $V_1$ term.  The first term on the right hand side of Eq.~(\ref{LinearResponse1}), is the so-called diamagnetic response, which involves only the ground-state density $n_0(\rv)$.
A formal inversion of Eq.~(\ref{LinearResponse1}) yields immediately an equation of motion for the current density:
\be\label{EOMFormal}
i \omega\int  d\rv' [\chi^{-1}]_{\mu\nu}(\rv,\rv',\omega)  \jv_{1,\nu}(\rv',\omega) =\partial_\mu V_1(\rv,\omega)\,.
\ee
Obviously, this result is purely formal, since we have no way to exactly calculate the current-current response function of a many-body system, let alone invert it. 
However Eq.~(\ref{EOMFormal}) can serve as a convenient starting point for constructing approximate theories. Following the ideas proposed in Refs.~\cite{PRLCM,CM} 
we consider the high frequency limit of Eq.~(\ref{EOMFormal}) and then interpret it as an approximate equation of motion for the induced current density.

We start by observing that, at high frequency,  the current-current response function has the well-known expansion~\cite{gvbook,keldysh} 
\ber\label{eomLR}
{\chi}_{\mu\nu}(\rv,\rv',\omega) &=& n_0(\rv) \delta(\rv-\rv')\delta_{\mu\nu} - i\frac{ {\cal B}_{\mu\nu}(\rv,\rv')}{ \omega} \nn\\
&+& \frac{M_{\mu\nu}(\rv,\rv')}{ \omega^2}+{\cal O}\left(\frac{1}{\omega^3}\right)\,,
\eer
where
\be\label{M1}
{\cal B}_{\mu\nu}(\rv,\rv') =  i\langle \psi_0 \vert [\hat j_\mu(\rv),\hat j_\nu(\rv')]\vert\psi_0\rangle\,,
\ee
and 
\be\label{M2}
M_{\mu\nu}(\rv,\rv') = - \langle \psi_0 \vert [[\hat H_0,\hat j_\mu(\rv)],\hat j_\nu(\rv')]\vert\psi_0\rangle\,.
\ee

A few words of caution should be added at this point. 
The expansion written for ${\bm \chi}$ in Eq.~(\ref{eomLR}) is done under the assumption
that the real-space current-current response function has a regular Taylor  expansion in inverse powers of $1/\omega$ at high-frequency.   However, this is not always the case. If the initial (ground) state is not sufficiently smooth, the unboundedness of the kinetic energy operator may cause $\chi_{\mu\nu}(\omega)$ to develop a non-analytic behavior (e.~g. fractional powers of $\omega$) at high frequency \cite{footnote}. Therefore, strictly speaking, the expansion of Eq.~(\ref{eomLR}) assumes that proper smoothness conditions are imposed on the initial state of the system. Fortunately, after inverting Eq.~(\ref{eomLR}) to get the high frequency expansion of the operator $\chi^{-1}_{\mu\nu}$ entering the left hand side of Eq.~(\ref{EOMFormal}), all smoothness restriction can be relaxed. It turns out that such obtained high frequency form  of the inverse response function is generally valid. We prove this in Secs.~IV and V by rederiving the equation of motion for the current density via the Lagrangian frame formalism, which can be viewed as a direct construction of a high frequency/short time limit of $\chi^{-1}_{\mu\nu}$. In Appendix B we explicitly demonstrate that for the one-particle system the above formal inversion procedure indeed yields the exact form of the inverse current response function. A possible nonanalytic behavior in $\omega$ of the response is correctly recovered in our theory because at the level of Eq.~(\ref{EOMFormal}) it is encoded in the space part of the operator $\chi^{-1}_{\mu\nu}$ acting on the current density.

Thus, we invert Eq.~(\ref{eomLR}) and plug the result into Eq.~(\ref{EOMFormal}).  This yields
\ber\label{eomLR3}
\omega^2 ~ j_{1,\mu}  &+&  i\omega
 \int d\rv' {\cal B}_{\mu\nu}(\rv,\rv') \frac{ j_{1,\nu}(\rv')}{n_0(\rv')} \nn \\
&-& \int d\rv' {\cal K}_{\mu\nu}(\rv,\rv') \frac{ j_{1,\nu}(\rv')}{n_0(\rv')} = \partial_\mu V_1
\eer
where
\ber\label{M2tilde}
 {\cal K}_{\mu\nu}(\rv,\rv') &=&  M_{\mu\nu}(\rv,\rv')  \nn \\ 
&+&  \int d\rv''  {\cal B}_{\mu\gamma}(\rv,\rv'') \frac{1}{n_0(\rv'')}  {\cal B}_{\gamma\nu}(\rv'',\rv').
\eer

Evaluation of the current-current commutator in Eq.~(\ref{M1}) is relatively straightforward and yields
\ber\label{M1bis}
{\cal B}_{\mu\nu}(\rv,\rv') &=&
j_{0,\mu}(\rv') \frac{\partial \delta(\rv - \rv')}{\partial r_\nu} +  j_{0,\nu}(\rv) \frac{\partial \delta(\rv' - \rv)}{\partial r_\mu} \nn\\
&+& \epsilon_{\mu \nu \gamma} B_{0,\gamma}(\rv) n_0(\rv) \delta(\rv -\rv') \,,
\eer
where $\jv_0(\rv)$ is the ground-state current.  Notice that ${\cal B}_{\mu\nu}(\rv,\rv')$ is imaginary and antisymmetric.

The evaluation of the double-commutator ${\cal K}_{\mu\nu}(\rv,\rv')$, while in principle equally straightforward, is in practice an extremely cumbersome task, leading to very complicated expression to which we are not able to attach any transparent physical meaning.  We do not undertake this task here, since in the next section we will develop an alternative approach, which leads more directly to a physically meaningful equation of motion.

Even without knowing the explicit form of ${\calKv}$  we can learn something from the form of Eq.~(\ref{eomLR3}).  To begin with, we may express the response of the current in terms of the displacement
defined in the conventional way, i.e. according to Eq.~(\ref{Displacement1.1}).   By doing so, we obtain
\be\label{eomDisp1}
\omega^2 ~ n_0 {\bf u}  + i\omega ~ \calBv \cdot {\bf u} - \calKv \cdot {\bf u} =  n_0 \nabla V_1\;,
\ee
where we have adopted a compact notation in which
\be\label{defO}
\calOv \cdot {\bf u} \equiv  \int d\rv' ~{\cal O}_{\mu\nu}(\rv,\rv')  u_{\nu}(\rv')\;.
\ee
It is easy to prove that  ${\bf M}$  must be a positive-definite operator, since, by definition, it is related to the second-order term in the expansion of the energy 
of the distorted ground-state~(\ref{DeformedPsi}) in powers of $\uv$. 

Setting the right-hand side to zero in Eq.~(\ref{eomDisp1}), we recognize a generalized eigenvalue problem of the form (for more details see Appendix A)
\be\label{eomDisp-B}
\omega^2 ~ n_0{\bf u}  + i \omega ~ {\calBv} \cdot {\bf u} - {\calKv} \cdot {\bf u} =  0\;.
\ee
Under the stronger assumption that not only ${{\bf M}}$, but also ${\calKv}$ is positive definite, 
one can show (see Appendix A) that the generalized eigenvalue problem in Eq.~(\ref{eomDisp-B}) has real solutions $\omega$ 
whose modes satisfy a generalized orthogonality relation
\be
\langle  {\uv}_B, i{\calBv} \cdot {\uv}_A \rangle ~+ \left( \omega_A + \omega_B \right) \langle {\uv}_B, n_0 {\uv}_A \rangle = 0\;,~~~ \omega_A \ne \omega_B
\ee
with the scalar product defined as follows
\be \label{product1}
\langle {\uv}_B,  {\uv}_A \rangle \equiv \int  d \rv ~  {u}^*_{B,\mu} (\rv)   {u}_{A,\mu}(\rv) \;.
\ee

The eigenvalues are naturally interpreted as (approximate) excitation energies and the corresponding eigenvectors are matrix elements of the current density operator between the ground-state and the excited state in question (see Ref. \onlinecite{CM} and Appendix B in the present paper).  Thus, we see that the assumption of positivity of ${\calKv}$ is, in practice, equivalent to the expectation that our approximation does not lead to spurious instabilities (i.e. complex excitation energies). 

A natural question at this point is: how reliable  our approximate equation of motion~(\ref{eomDisp1}) will be in situations other than the very high frequency limit?  The answer to this question has already been discussed
extensively in Ref. \onlinecite{CM}, so we only summarize the main points adding the required modifications where needed.

First of all, the fact that the proposed current-current response function has the exact high-frequency behavior up to order $1/\omega^2$ implies that the first moment of the spectral function $\int_0^\infty d\omega ~ \omega \Im m \chi_{\mu\nu}(\rv,\rv',\omega)  \simeq M_{\mu\nu}(\rv,\rv')$ is exactly reproduced.  
Thus, while the energy of individual excitations may be misrepresented, the first moment of the current-density fluctuation excitation spectrum (or, equivalently, the third moment of the density fluctuation excitation spectrum) is correctly reproduced.

Second, our equation assumes an exact and explicit form for one-particle systems.
This is rigorously proved in Appendix B and Appendix C. In particular, in Appendix B 
we show that ${\bm \chi}^{-1}(\omega) \cdot \jv_1 =  A+iB/\omega+C/\omega^2$ for {\em any} frequency $\omega$,
where $A$, $B$, and $C$ are coefficients expressible in terms of the ground-state density and current density.
The coefficient $A$ is related to the diamagnetic response, while $B$ and $C$ may be related to the coefficients of $\omega^{-1}$ 
and $\omega^{-2}$  in the high-frequency expansion of ${\bm \chi}$.

The fact that the form derived by the above procedure becomes exact in one-particle systems has a deeper physical significance.  
Recall that in Refs.~\onlinecite{PRLCM,CM} the equation for the displacement was derived by describing the dynamics in a {\em co-moving } reference frame, 
defined as an non-inertial frame in which the density remains constant and the current density is always zero.   So the high-frequency approximation could be restated in terms of an
 ``anti-adiabatic approximation" or ``elastic approximation"  introduced as the assumption that not only the density and the current, but the wave function itself remains 
 unchanged in the co-moving reference frame.  
 This assumption is correct for one-particle systems, because in such systems the density and the current density  uniquely determine the wave function, up to a trivial phase factor. 

We will show that  this approach --  transformation to a co-moving frame and the requirement of a stationary wave function in that frame -- also works in the presence of a magnetic field, but with an important difference: the current density in the co-moving frame will not be zero, but will be equal to the current density in the ground-state.  Because of this redefinition the relation between current response and displacement field will be drastically changed: Eq.~(\ref{Displacement1}) will be replaced by Eq.~(\ref{Displacement2}).
However, as a result of this modification, we will be able to obtain an explicit equation of motion for the new $\uv$, bypassing the need to calculate cumbersome commutators.  The final form of the equation of motion will be a natural extension of the one derived in
Refs.~\onlinecite{PRLCM,CM} in the absence of a magnetic field.

\section{Lagrangian formulation}\label{III}

The key physical idea of the Lagrangian formulation in application to many-body dynamics is to separate the convective motion of the electron fluid from the motion of electrons relative to the convective flow. The former type of motion is characterized by the trajectories of infinitesimal volume elements, while the latter is described by a many-body Schr\"odinger equation in a local non-inertial reference frame attached to those elements \cite{TokatlyPRB2007}. 

In the standard Lagrangian formalism,~\cite{TokatlyPRB2007} the convective dynamics of the system is described by a  set of trajectories $\rv(\xiv,t)$, which represent the motion of an infinitesimal volume elements initially located at $\xiv$.   These trajectories satisfy the  first-order differential equation
\be
\partial_t \rv(\xiv,t)= \frac{\jv(\rv(\xiv,t),t)}{n(\rv(\xiv,t),t)}  
\ee
with initial condition $\rv(\xiv,0)=\xiv$.   Notice that we are not even making the linear response approximation at this stage: the quantity $n(\rv(\xiv,t),t)$ is the actual particle density along the trajectory and reduces to the ground-state density $n_0(\xiv)$ only at the initial time.   The displacement field $\uv(\xiv,t)$ is defined via the relation
$\rv(\xiv,t)=\xiv+\uv(\xiv,t)$.

This standard definition is, however, inconvenient if the initial stationary/ground state already carries a non-zero current-density $j_{0}^{\mu}=n_0v_{0}^{\mu}$, which is the case in a magnetized electronic system.  Indeed, even in the absence of the external driving field the ground state flow drags the volume elements and produces a time-dependent displacement in a system that, physically, should be considered as stationary and undeformed.  It is natural to try and exclude the equilibrium convective motion from the trajectory function, i.e., to consider a volume element as moving only when it moves {\it relatively} to the ground-state flow. To accomplish this in the presence of the magnetic field we adopt a modified Lagrangian description, where the motion of the volume elements is described relatively to the ground state flow.  Thus, when the system is in the ground-state the displacement vanishes and the volume elements are regarded as stationary.  An observer ``riding on"  these volume elements does not move at all (relatively to the laboratory) and sees the current density $\jv_0$.  Now, when the system is out of equilibrium a finite displacement appears and the volume elements begin to move.  However, to an observer attached to the material elements the current density still appears to be $\jv_0$.    {\it Just as in the standard Lagrangian description an observer riding on the volume element sees stationary density $n_0$ and zero current density, in the present description an observer riding on the volume element sees stationary density $n_0$ and current density $\jv_0$.} 

The task at hand is now to implement the transformation that makes the density and the current density equal to their ground-state values at all times.  This is done in two steps.  First we learn how to describe the quantum many-body dynamics in a generic non-inertial reference frame.  Then we fix the reference frame from the condition that density and current density retain their ground-state values at all times.  We carry out the above steps without assuming that the displacement is small, i.e., in a fully non linear way.  The linearization of the equations of motion will be carried out only at the very end of the next Section.

\subsection{Transformation to a non-inertial reference frame}

Let us consider a local reference frame moving along a prescribed
trajectory  $\rv(\xiv,t)$.  The form of the quantum many-body dynamics in such a frame 
is worked out in full detail in Refs.~\onlinecite{TokatlyPRB2005a,TokatlyPRB2005b,TokatlyPRB2007}. Here we present only the key results that are needed for our  purposes. 

First of all,  the transformed many-body wave function has the form
\begin{eqnarray}
\widetilde{\Psi}({\xiv}_{1},\dots,{\xiv}_{N},t)
&=& \prod_{j = 1}^{N}g^{\frac{1}{4}}({\xiv}_{j},t)  e^{-iS_{\text{cl}}({\xiv}_{j},t)} \nn \\ &\times&
\Psi({\bf r}({\xiv}_{1},t),\dots,{\bf r}({\xiv}_{N},t),t)\,,
\label{PsiLagr}
\end{eqnarray}
where
\be\label{MetricTensor}
g_{\mu\nu}(\xiv,t) = \frac{\partial r^{\alpha}(\xiv,t)}{\partial\xi^{\mu}}
\frac{\partial r^{\alpha}(\xiv,t)}{\partial\xi^{\nu}}
\ee
is the metric tensor induced by the (non-singular) transformation $\rv \rightarrow \xiv$,
$\sqrt{g} \equiv \sqrt{\det g_{\mu\nu}}$ is the Jacobian of the same transformation
and
\begin{eqnarray}
\label{action}
S_{\text{cl}}({\bm\xi},t) &=& \int_0^t d t \left[ 
\frac{1}{2}(\dot{\bf r}({\bm\xi},t))^{2}  - \dot{\bf r}({\bm\xi},t){\bf A}({\bf r}({\bm\xi},t),t)  \right. \nn \\ 
&-&  \left. V({\bf r}({\bm\xi},t),t)\right].
\end{eqnarray}
is the classical action of a particle moving along the assigned
trajectory.  The factor $\prod_{j = 1}^{N}g^{\frac{1}{4}}({\bm\xi}_{j},t)$ in Eq.~(\ref{PsiLagr})  preserves the standard normalization of the wave function $\langle\widetilde{\Psi}|\widetilde{\Psi}\rangle=1$ after a non-volume-preserving transformation of coordinates. Equation (\ref{PsiLagr}) is a generalization of the transformation to a homogeneously accelerated frame, which is used, for example, in the proofs of a harmonic potential theorem \cite{Dobson1994,Vignale1995a,Baer}.

The transformed wave function $\widetilde{\Psi}({\bm\xi}_{1},\dots,{\bm\xi}_{N},t)$ satisfies the Schr\"odinger equation
\begin{equation}
 \label{SELagr}
i\partial_t\widetilde{\Psi}({\bm\xi}_{1},\dots,{\bm\xi}_{N},t) =
\widetilde{H}[g_{\mu \nu},\bm{\mathcal A}]\widetilde{\Psi}({\bm\xi}_{1},\dots,{\bm\xi}_{N},t)
\end{equation}
with the transformed Hamiltonian 
\begin{equation}
 \label{HLagr}
\widetilde{H}[g_{\mu \nu},\bm{\mathcal A}] 
= \sum_{j = 1}^{N}g^{-\frac{1}{4}}_{j}
\hat{K}_{j,\mu}\frac{\sqrt{g_{j}}g^{\mu\nu}_{j}}{2}
\hat{K}_{j,\nu}g^{-\frac{1}{4}}_{j}
+\frac{1}{2}\sum_{k\ne j}W(l_{\bm\xi_{k}\bm\xi_{j}})
\end{equation}
where $\hat{K}_{j,\mu}=-i\partial_{\xi^{\mu}_{j}}
+ {\cal A}_{\mu}(\bm\xi_{j},t)$,
\ber
 \label{Aeff}
{\mathcal A}_{\mu}(\xiv,t) &=& \frac{\partial r^{\nu}}{\partial\xi^{\mu}}
A_{0,\nu}({\bf r}(\bm\xi,t))
-  \frac{\partial r^{\nu}}{\partial\xi^{\mu}}
\dot{r}^{\nu}({\bf r}(\bm\xi,t),t) \nn \\
&+& \partial_{\xi^{\mu}}S_{\text{cl}}({\bm\xi},t)\,,
\eer
and  $l_{\bm\xi_{k}\bm\xi_{j}}$  is the distance between $j$th and $k$th particles in the non-inertial frame (i.e., the length of geodesic 
connecting points $\bm\xi_{j}$ and $\bm\xi_{k}$ in the space with metric $g_{\mu\nu}$). 
The metric tensor $g_{\mu,\nu}$ and ``effective" vector potential ${\mathcal A}$ in Eq.~(\ref{Aeff}) describe the combined action of  external forces and inertial forces in the local non-inertial frame.

The transformed densities are obtained from the reduced one-particle density matrix
\ber\label{tilde-gamma}
\widetilde{\gamma}(\xiv,\xiv',t) = N &\int& \prod_{j =2}^{N}d\xiv_{j} \nn \\
&\times& \widetilde{\Psi}^*(\xiv,\dots,\xiv_{N},t)\widetilde{\Psi}(\xiv',\dots,\xiv_{N},t)\;
\eer
through the formulas:
\be\label{tilde-n}
\widetilde{n}(\xiv,t) = \widetilde{\gamma}(\xiv,\xiv,t),
\ee
\ber\label{tilde-j}
\widetilde{j}^{\mu}(\xiv,t)  = g^{\mu \nu}(\xiv,t)  \left[  \widetilde{j}_{p,\nu }(\xiv,t) + \widetilde{n}(\xiv,t) ~{\cal A}_\nu(\xiv,t)  \right]\;,
\eer
where 
\be
\widetilde{j}_{p,\mu}(\xiv,t) =  \frac{i}{2}\lim_{{\xi^{\mu}}' \to\xi^{\mu}}(\partial_{\xi^\mu} -\partial_{{\xi^{\mu}}'})\widetilde{\gamma}(\xiv,\xiv',t)\;, 
\ee
is the (covariant) paramagnetic current.  It is extremely important to carefully keep track of covariant (lower indices) and contravariant (upper indices) components of vectors and tensors.  The two types of components are connected by the metric tensor via the standard formulas
\be
V_\mu = g_{\mu\nu}V^\nu\,~~~~~V^\mu = g^{\mu\nu}V_\nu\,,
\ee
where $g^{\mu\nu}$ is the inverse of $g_{\mu\nu}$.
Armed with these definitions, one can readily verify that the expressions of the densities in the non-inertial frame are related to the ones in the laboratory frame
by  the relations
\begin{eqnarray}
\label{tilde-n2}
\tilde{n}(\xiv,t) &=& \sqrt{g}n(\xv(\xiv,t),t),\\
\label{tilde-j2}
\tilde{j}^{\nu}(\xiv,t) &=& \sqrt{g}\frac{\partial\xi^{\nu}}{\partial x^{\mu}}
\left[j^{\mu}(\xv(\xiv,t),t)  \right. \nn \\
 &-& \left. n(\xv(\xiv,t),t)\dot{r}^{\mu}(\xv(\xiv,t),t)\right]\;.
\end{eqnarray}

The local conservation laws for the transformed many-particle problem read as follows 
\be\label{t-continuity}
\partial_t \widetilde{n} + \partial_{\xi^\mu} \widetilde{j}^\mu = 0 \;
\ee
and 
\begin{equation}
 \label{NSg}
\partial_{t}\widetilde{j}_{\mu} - \widetilde{j}^{\nu}
(\partial_{\xi^{\nu}}{\cal A}_{\mu} - \partial_{\xi^{\mu}}{\cal A}_{\nu})
- \widetilde{n}~\partial_{t}{\cal A}_{\mu}
+ \sqrt{g}~D_{\nu}\widetilde{P}^{\nu}_{\mu} = 0\,,
\end{equation}
where $\widetilde{P}^{\nu}_{\mu}$ is a mixed component of the stress tensor, which, in 
Ref.~\onlinecite{TokatlyPRB2007} was proved to have the form
\be
\label{StressLagr}
\widetilde{P}^{\mu\nu}(\bm\xi,t) = - \frac{2}{\sqrt{g}}\left\langle\widetilde{\Psi}\left\vert
\frac{\delta\widetilde{H}[g_{\alpha\beta},
\bm{\mathcal{A}}]}{\delta g_{\mu\nu}(\bm\xi,t)}\right\vert\widetilde{\Psi}\right\rangle\;, 
\ee
where the functional derivative with respect to $g_{\mu\nu}$ is to be taken fixed $\bm{\mathcal{A}}$.  

The quantity $D_{\nu} \widetilde{P}^{\nu}_{\mu}$ in Eq.~(\ref{StressLagr}) stands for the {\it covariant divergence} of the stress tensor and is explicitly given by  
\be
D_{\nu} \widetilde{P}^{\nu}_{\mu}=\frac{1}{\sqrt{g}}\partial_\nu \sqrt{g} \widetilde{P}^{\nu}_{\mu}-\frac{1}{2}\widetilde{P}^{\alpha\beta}\partial_\mu g_{\alpha\beta}\,.
\ee

A rather lengthy calculation shows that
\be\label{StreeRel}
\sqrt{g}D_{\nu}\widetilde{P}^{\nu}_{\mu} 
= \frac{\partial{r^\nu}}{\partial {\xi^\mu}} \left\langle \widetilde{\Psi} \left\vert \left. ~ \left( \frac{\delta \widetilde{H}}{\delta r^\nu}\right) \right\vert_{\bm{\mathcal{A}}} ~ \right\vert \widetilde{\Psi} \right \rangle\;,
\ee
where the functional derivative with respect to $r^\nu$ is to be taken at constant ${\bm{\mathcal{A}}}$. It is important to notice that in the above equation, the transformed hamiltonian $\tilde H$ -- a functional of $g_{\mu\nu}$ and ${\cal A}$ [see Eq.~(\ref{HLagr})] --  is actually treated as a functional of the trajectory $r_\mu$ and ${\bm{\mathcal{A}}}$.  This is permissible because $g_{\mu\nu}$ itself is a functional of $r_\mu$ -- see Eq.~(\ref{MetricTensor}).

We will make heavy use of this identity in the next section.  Now let us learn how to fix our reference frame so that the density and current density become constants of the motion.

\subsection{Fixing the reference frame}

As discussed in the introduction to this section, our goal is to separate the convective motion of the electron fluid from the motion of electrons relative to the convective flow. 
This is  achieved by following the  volume elements along their trajectories, so that
the density and the current density becomes stationary, with values equal to the initial one.
Mathematically, this translates to the single condition
\be\label{fix-j}
\widetilde{j}^{\mu}(\xiv,t) = \widetilde{j}^{\mu}(\xiv,0) = j^{\mu}_0(\xiv)\;.
\ee

It is important to notice that our condition must be imposed on the {\it contravariant} components $\widetilde{j}^{\mu}$ of the current density, rather than on the covariant components $\widetilde{j}_{\mu}$.  The two choices are, at first sight, equally plausible, but not equivalent, since the metric tensor $g_{\mu\nu}$, which connects covariant and contravariant components of vectors, is time-dependent. 
What forces our choice is the fact that the contravariant component of the current density  appears naturally in the continuity equation~(\ref{t-continuity}).     
Thus, with this choice, the continuity equation  (\ref{t-continuity}) guarantees that we get
\be\label{fix-n}
\widetilde{n}(\xiv,t) = \widetilde{n}(\xiv,t=0) = n_0(\xiv)\;
\ee
as soon as Eq.~(\ref{fix-j}) is satisfied.

Given Eq.~(\ref{fix-j}) and Eq.~(\ref{fix-n}), 
the effective potential acting on the system is determined through Eq.~(\ref{tilde-j}) as follows
\ber\label{Aeff3}
{\cal A}_\mu(\xiv,t) = g_{\mu \nu}(\xiv,t)  v_0^{\nu}(\xiv) - \frac{ \widetilde{j}_{p,\mu}(\xiv,t) }{n_0(\xiv) }\;,
\eer
where $v_0^{\mu}(\xiv) = j_0^\mu(\xiv)  / n_0(\xiv) $\;.
Eq.~(\ref{Aeff3}) is a gauge-fixing condition which determines the effective vector potential $\bm{\mathcal{A}}$ and, thus, a particular  non-inertial frame.

\subsection{Equation of motion for the Lagrangian trajectory}

After fixing the reference frame the system is completely characterized by two dynamical variables -- the trajectory of volume elements $r_\mu(\xiv,t)$, and the transformed many-body wave function $\widetilde{\Psi}(t)$ that describes microscopic motion relative to the convective flow. The wave function $\widetilde{\Psi}(t)$ satisfies the Schr\"odinger equation (\ref{SELagr}).
In order to obtain a physicaly trasparent form of the equation of motion for the Lagrangian trajectory $r_\mu(\xiv,t)$ we differentiate Eq.~(\ref{Aeff}) with respect to time and we get
\begin{equation}
 \label{traj1}
\ddot{r}^{\mu}  + \partial_{\mu}V_0(\rv) + [\dot\rv\times{\bf B_0}(\rv)]_{\mu} 
+ \frac{\partial \xi^{\nu}}{\partial r^{\mu}}
\partial_{t}{\cal A}_{\nu} = - \partial_{\mu}V_1(\rv,t)\;.
\end{equation}

In this equation we insert the value of  $\partial_{t}{\cal A}_{\nu}$ determined through the local momentum balance equation [Eq.~(\ref{NSg})], i.e.,
\ber\label{NSg2}
n_0\partial_{t}{\cal A}_{\mu} &=&  \partial_{t} \left( g_{\mu \nu}j^{\nu}_{0} \right)  - j^{\nu}_{0}
(\partial_{\xi^{\nu}}{\cal A}_{\mu} - \partial_{\xi^{\mu}}{\cal A}_{\nu}) \nn \\ 
&+&\sqrt{g}D_{\nu}\widetilde{P}^{\nu}_{\mu}\;,
\eer
where we have made use of the conditions ~(\ref{fix-j}) and~(\ref{fix-n}) and lowered the index of the contravariant current $\widetilde{j}_{\mu} = g_{\mu \nu} \widetilde{j}^{\nu}$ through the action of the metric tensor $g_{\mu \nu}$.
And now,
on the right hand side of Eq.~(\ref{NSg2}) we plug in the expression ~(\ref{Aeff}) for ${\cal{A}}_\mu$  in the combination $(\partial_{\xi^{\nu}}{\cal A}_{\mu} - \partial_{\xi^{\mu}}{\cal A}_{\nu}) $ (notice that the contribution of the classical action term vanishes, since the curl of a gradient is zero), and make use of the identity~(\ref{StreeRel}) for $ \sqrt{g}D_{\nu}\widetilde{P}^{\nu}_{\mu}$.  The result of these manipulations is
\ber\label{intermediate}
&&\ddot{r}^\mu + 2 v_0^\nu \partial_\nu \dot{r}^\mu + \left[ \left( D_t \rv \right) \times \Bv_0 \right]_\mu + \partial_\mu V_0(\rv) \nn\\
&&+ \frac{1}{n_0}\left \langle \widetilde{\Psi} \left\vert \left. ~ \left( \frac{\delta \widetilde{H} }{\delta r^\nu}\right)\right\vert_{\bm{\mathcal{A}}} ~ \right\vert\widetilde{\Psi} \right\rangle = -\partial_\mu V_1(\rv,t)\;,
\eer
where 
\be\label{conv0}
D_t = \partial_t + v_0^\nu \partial_{\nu}\;
\ee
is a ``convective time derivative", which takes into account the ground-state flow.

In view of the last result, it is desirable to change {\it all} the time derivatives  in Eq.~(\ref{intermediate})  to convective ones. 
To this end, we note that
\be
 \ddot{r}^\mu + 2 v_0^\nu \partial_\nu \dot{r}^\mu = \left(\partial_t^2+2 v_0^\nu\partial_\nu \partial_t\right)r^\mu
 \ee
and ``complete the square" to get
\ber\label{CompleteSquare}
 \ddot{r}^\mu + 2 v_0^\nu \partial_\nu \dot{r}^\mu &=& D_t^2r^\mu- v_0^\alpha\partial_\alpha v_0^\beta\partial_\beta r^\mu\nn\\
&=&D_t^2r^\mu+ \frac{1}{n_0} \frac{\delta W_{0}[g_{\mu \nu}]}{\delta r^\mu}\,, 
 \eer
where
\be\label{W0}
W_{0}[g_{\mu \nu}] \equiv  \frac{1}{2}  \int d \xiv ~ n_0(\xiv) g_{\mu \nu}(\xiv,t) v_0^{\mu}(\xiv) v_{0}^{\nu}(\xiv)\;.
\ee
Using Eq.~(\ref{CompleteSquare}) in Eq.~(\ref{intermediate}), we  obtain
\ber\label{trajectory}
&&D^2_t r^\mu  + \left[ \left( D_t \rv \right) \times \Bv_0 \right]_\mu + \partial_\mu V_0(\rv) \nn\\&&+
\frac{1}{n_0} \left\{ \left\langle \widetilde{\Psi}\left\vert ~ \left. \left( \frac{\delta \widetilde{H} }{\delta r^\mu}  \right)  \right\vert_{\bm{\mathcal{A}}}\right\vert\widetilde{\Psi} ~ \right\rangle  \right. + \left. \frac{\delta W_{0} }{\delta r^\mu} \right\} = -\partial_\mu V_1(\rv,t)\nn\\ \;.
\eer

We stress that the latter equation, Eq.~(\ref{trajectory}), is an {\em exact nonlinear equation} for the trajectory of current-carrying material elements in the quantum many-body system, 
provided $\widetilde{\Psi}(t)$ entering the stress force is the solution to the Schr\"odinger equation (\ref{SELagr}).

\section{Elastic approximation}\label{IV}

We now introduce our {\em elastic approximation} by setting
\be\label{Psi0}
\widetilde{\Psi}(\xiv,t) \equiv \Psi_0(\xiv)\;,
\ee
It is clear that this approximation is consistent with the requirement of stationary transformed densities, but cannot be exact in general, since the correct form of $\widetilde\Psi$ is determined by Eq.~(\ref{SELagr}).

The approximation in Eq.~(\ref{Psi0}) applies to {\em short-time} dynamics, or to {\em fast-driving} fields such that there is no time for the
wave function to  adjust to minimize the energy in the presence of fast-varying external conditions. Here, we may recognize the {\em high-frequency approximation} discussed in Section~\ref{II}.  
The same approximation may be characterized as an  {\em anti-adiabatic approximation} in the following sense. 
The wave function in the Eulerian (inertial) frame is obtained as an instantaneous deformation of the initial wave function.   While the initial wave function minimizes the energy (i.e., it is the ground-state), it is clear that the deformed wave function does not minimize, at any given instant, the energy of the system.  The deformation is
similar  to the change in the shape of an elastic body. Hence, the term {\em  elastic approximation}.

Let us now replace Eq.~(\ref{Psi0}) into the trajectory equation~(\ref{trajectory}).
Because the ground-state wave function $\Psi_0$ does not depend on the Lagrangian trajectories, we can pull the functional derivative with respect to $r^\mu$ out of the quantum average.  
In other words, we can write 
\be\label{trajectory4}
\left. \left\langle \widetilde{\Psi} \left\vert ~ \left(  \frac{\delta \widetilde{H}}{\delta r^\mu}   \right) \right\vert_{ \bm{\mathcal{A}}} ~ \right\vert\widetilde{\Psi} \right\rangle
+\frac{\delta W_{0}}{\delta r^\mu} =  \left.\frac{\delta }{\delta r^\mu} \langle \Psi_0 |  \widetilde{H}+\hat W_0 | \Psi_0 \rangle  \right\vert_{ \bm{\mathcal{A}} }
\ee
where $\hat W_0$ is the operator
\be\label{W0HAT}
\hat W_{0}[g_{\mu \nu}] \equiv  \frac{1}{2}  \int d \xiv ~\hat n (\xiv) g_{\mu \nu}(\xiv,t) v_0^{\mu}(\xiv) v_{0}^{\nu}(\xiv)\;,
\ee 
such that $\langle \psi_0|\hat W_0|\psi_0\rangle = W_0$.

The functional derivative with respect to $r^\mu$ at constant $ \bm{\mathcal{A} }$ can be written as an unrestricted functional derivative minus counter-terms, 
which cancel the unwanted contributions from the variation of $ \bm{\mathcal{A}}$. In formulas, we have
\ber\label{DerivativeFiddling}
&& \left.\frac{\delta}{\delta r^\mu(\xiv)}   \langle \Psi_0 |   \widetilde{H}+\hat W_0  | \Psi_0 \rangle \right\vert_{ \bm{\mathcal{A}} }= 
\frac{\delta}{\delta r^\mu(\xiv)} \langle  \Psi_0 |   \widetilde{H}+\hat W_0  | \Psi_0 \rangle
\nn \\  
&-& \int d\xiv' \left.\frac{\delta {\cal A}_\nu(\xiv')}{\delta r^\mu(\xiv)} \left( \frac{\delta}{\delta {\cal A}_\nu(\xiv')}
 \langle \Psi_0 |   \widetilde{H}  | \Psi_0 \rangle\right\vert_{\rv}\right)\;,
\eer
where we have used the fact that $\hat W_0$ does not depend on ${ \bm{\mathcal{A}} }$.
Now we recall the identity
\be
\hat{ \widetilde{j}}^\mu(\xiv')= \frac{\delta \tilde H}{\delta {\cal A}_\mu(\xiv')}\;
\ee
and the fact that  $\hat{ \widetilde{j}}^\mu$ as well as ${ \cal A }_\nu$ [see Eq.~(\ref{Aeff3})] must be here evaluated at $\Psi_0$.
In particular, we notice that, according to Eq.~(\ref{Aeff3}) 
\be
\frac{\delta {\cal A}_\nu(\xiv')}{\delta r^\mu(\xiv) } =  \frac{  \delta g_{\nu \alpha}(\xiv',t)  v_0^{\alpha}(\xiv') }{\delta r^\mu(\xiv) }\;.
\ee 
Therefore, Eq.~(\ref{DerivativeFiddling}) can be compactly and suggestively restated as
\ber\label{ElasticForce}
\left.\frac{\delta}{\delta r^\mu(\xiv)}   \langle \Psi_0 | \widetilde{H} + \hat W_0 | \Psi_0 \rangle \right\vert_{ \bm{\mathcal{A}} }&=&\frac{\delta E_{el}[g_{\mu\nu}]}{\delta r^\mu(\xiv)}\nn\\
&\equiv&- {\cal F}_{{\rm el},\mu}(\xiv)\,.
\eer 
Here $E_{\rm el}[g_{\mu\nu}]$ is an elastic energy defined as
\be
E_{\rm el}[g_{\mu\nu}]=\langle \psi_0| \tilde H[g_{\mu\nu},{\cal A}= -\vv_{p0}]|\psi_0\rangle\,,
\ee
where  $\vv_{p0}=\jv_{p0}/n_0$ is the paramagnetic velocity in the ground-state (in a proper gauge, this quantity always vanishes in a one-particle system).
We shall refer to ${\fbold}_{\rm el}$, as  the elastic force acting on the systems: this is given by the functional derivative of the elastic energy with respect to the Lagrangian trajectory.

Plugging the above definitions and formulas into the trajectory equation~(\ref{trajectory}),  we finally obtain
\be\label{trajectory5}
D^2_t r^\mu  + \left[ \left( D_t \rv \right) \times \Bv_0 \right]_\mu + \partial_\mu V_0 -  \frac{1}{n_0}{\cal F}_{{\rm el},\mu}   = -\partial_\mu V_1 \;.
\ee
The situation is actually analogous to what
happens in the case of a classical spring: displace it from the equilibrium, compute the gain in energy, get the elastic force from the derivative 
of the energy with respect to displacement. Classical elasticity theory generalizes this  to continuum media,  for which the ``stress field" is the functional derivative of the energy  with respect to the strain field.~\cite{CET}  Here, in a quantum many-body system, the corresponding quantity is the functional derivative of the (transformed) quantum energy with respect to the trajectories of the infinitesimal volume elements.   
Due to the presence of an external static magnetic field, those material elements carry a non-vanishing current. 

The quantum elastic energy can be naturally split into a kinetic component and an electron-electron interaction component. 
In detail, we find
\be
\label{E-el-magn1}
E_{{\rm el}}[g_{\mu\nu}]  =T_{{\rm el}}[g_{\mu\nu}] + W_{{\rm el}}[g_{\mu\nu}]\;,
\ee 
where
\begin{eqnarray}
\label{Ekin-magn1}
T_{{\rm el}}  [g_{\mu\nu}]  &=& \int d\bm\xi \left[\frac{\sqrt{g}g^{\mu\nu}}{2}\left(\partial_{\mu}\sqrt{\frac{n_0}{g^{1/2}}} \right)\left(\partial_{\nu}\sqrt{\frac{n_0}{g^{1/2}}}\right) \right. \nn \\
&+&\left. \frac{1}{2}\Delta T_{\mu\nu}^{(0)}g^{\mu\nu}\right]\;,
\eer
\begin{eqnarray}
\label{Delta-T-magn1}
\Delta T_{\mu\nu}^{(0)} &=& \frac{1}{2}\lim_{\xi'\to\xi}(\hat\pi^{*}_{\mu}\hat\pi'_{\nu} + \hat\pi^{*}_{\nu}\hat\pi'_{\mu})\gamma_0(\xiv,\xiv') \nn \\
&-&
(\partial_{\mu}\sqrt{n_0})(\partial_{\nu}\sqrt{n_0})\;,
\end{eqnarray}
$\gamma_0(\rv,\rv')$ is the ground-state one-particle density matrix
\ber\label{gamma0}
\gamma_0(\xiv,\xiv') = N &\int& \prod_{j =2}^{N}d\rv_{j} \nn \\
&\times& {\Psi}_0^*(\xiv,\dots,\xiv_{N})\Psi_0(\xiv',\dots,\xiv_{N})\;,
\eer
$\hat{\bm\pi}$ is the operator of the kinematic momentum for the relative motion 
\begin{equation}
\label{pi}
\hat{\bm\pi} = -i\nabla - \vv_{0p} \;,
\end{equation}
\ber
\label{Eint-magn1}
W_{{\rm el}}  [g_{\mu\nu}]  &=& \frac{1}{2}\int d\bm\xi d\bm\xi' ~W({\bf r}(\bm\xi)-{\bf r}(\bm\xi'))\Gamma_0(\bm\xi,\bm\xi')\,.\nn\\
\end{eqnarray}
and 
\begin{eqnarray}
\label{Gamma0}
\Gamma_0(\rv,\rv') &=& N(N-1)\int \prod_{j =3}^{N}d\rv_{j} \nn \\
&\times&
\Psi_0^*(\xiv,\xiv',\dots,\xiv_{N})\Psi_0(\xiv,\xiv',\dots,\xiv_{N})
\end{eqnarray}
is the ground-state pair distribution function.  

The identification of the elastic energy defined by Eqs.~(\ref{E-el-magn1}-\ref{Gamma0}) is the key result of this section. 
In comparison with the non-magnetic case,\cite{PRLCM,CM} the essential difference is  the redefinition of the  kinetic contribution 
[see Eq.~(\ref{Ekin-magn1}), Eq.~(\ref{Delta-T-magn1}), and Eq.~(\ref{pi})].

\section{Linearized equation of motion}~\label{V}

The equation of motion for the Lagrangian trajectory derived in the previous section are fully nonlinear.  Here we present the linearized form of those equations, which are expected to be useful for systems performing small oscillations about the ground-state.

First of all, we insert Eqs.~(\ref{fix-j}) and ~(\ref{fix-n}) on the left hand sides of Eqs.~(\ref{tilde-n}) and~(\ref{tilde-j}).
This gives
\begin{eqnarray}
\label{tilde-n2}
n_0(\xiv) &=& \sqrt{g}n(\xv(\xiv,t),t),\\
\label{tilde-j2}
j^{\nu}_0(\xiv) &=& \sqrt{g}\frac{\partial\xi^{\nu}}{\partial x^{\mu}}
\left[j^{\mu}(\xv(\xiv,t),t)  \right. \nn \\
 &-& \left. n(\xv(\xiv,t),t)v^{\mu}(\xv(\xiv,t),t)\right]\;,
\end{eqnarray}
where
\be\label{r-disp}
\rv(\xiv,t) = \xiv + \uv(\xiv,t)\;.
\ee
Expanding to first order in $\uv$ we get
\be \label{n-u}
n(\xiv,t) = n_0(\xiv) - \nabla \cdot \left[ n_0(\xiv) \uv(\xiv) \right]\; \\
\ee
and
\be\label{disp-mag}
\jv(\xiv,t) = \jv_{0}(\xiv) + n_0(\xiv) \uv(\xiv) + \nabla \times \left[  \uv(\xiv) \times \jv_0(\xiv)   \right]\;.
\ee
Eq.~(\ref{disp-mag}) expresses the response of the current density, $\jv-\jv_0$, as the sum of two terms: the first accounts for  the polarization and the second for the {\em orbital} magnetization of the quantum medium.
We have thus succeeded in deriving the relation~(\ref{Displacement2}) between the linearized response of the current and the displacement field 
that was put forward in the introduction on a heuristic basis. 

We notice that, up to the linear order, there is no difference between the Lagrangian and the Eulerian description  of the displacement field, thus $\uv(\xiv) = \uv(\rv)$. 
In this spirit, we replace $\xiv$ by $\rv$ is all the expressions below. 
It is also worth emphasizing that Eqs.~(\ref{n-u}) and (\ref{disp-mag}) are general and hold true independently of the elastic or any other approximation.

The linearized equation of motion of the displacement is readily obtained 
\ber\label{eom.1}
D_t^2 \uv(\rv,t)  &+& \left[ D_t \uv (\rv,t) \right] \times \Bv_0(\rv)  + \left( \uv\cdot\nabla \right) \partial_\mu {\it V}_{\rm 0}(\rv) \nn\\ 
&+& \vv_0 \times  \left( \uv \cdot  \nabla \right) \Bv_{0}(\rv)  -  \frac{1}{n_0(\rv)} \fbold_{\rm el}(\rv,t)  \nn \\
&=& -\nabla V_1(\rv,t)\,.
\eer
where $\fbold_{\rm el}(\rv,t)$ is the (linearized) elastic force.  The terms on the second line result from the expansion of the classical forces to first order in $\uv$.
The linearized elastic force is given by
\be\label{fv1}
{\cal F}_{{\rm el}, \mu}(\rv,t) =-\int d\rv' \left.\frac{\delta^2 E_{{\rm el}} [\uv]}{\delta u_\mu (\rv)\delta u_\nu (\rv')}\right \vert_{\uv=0} u_\nu(\rv',t)\,,
\ee
and can be naturally separated into kinetic and electron-electron interaction contributions:
\be
{\cal F}_{{\rm el},\mu}(\rv) ={\cal F}^{\rm kin}_{{\rm el},\mu}(\rv) +{\cal F}^{\rm int}_{{\rm el},\mu}(\rv)\,. 
\ee

 Explicit expressions for the two components are obtained by closely following the steps outlined in Ref.~\onlinecite{PRLCM}.  The final expressions are 
\ber\label{Fk}
&&{\cal F}^{\rm kin}_{{\rm el},\mu}(\rv) =
\partial_\alpha[{\rm 2}{\it T}^{\rm mag}_{\nu\mu,0}(\rv){\it u}_{\nu\alpha}(\rv) + {\it T}^{\rm mag}_{\nu\alpha,0}(\rv)\partial_\mu {\it u}_\nu(\rv)]
\nn\\
&-&\frac{1}{4}\partial_\nu\partial_\mu \left[ n_0(\rv)\partial_\nu \nablabold\cdot\uv(\rv) \right] \nn\\
&+&\frac{1}{4}\partial_\nu \left\{2 \left[ \nabla^2 n_0(\rv) \right] u_{\nu\mu}(\rv) +\left[ \partial_\nu n_0(\rv) \right] \partial_\mu \nablabold\cdot\uv (\rv) \right.\nn\\
&+&\left. \left[\partial_\mu n_0(\rv)\right] \partial_\nu \nablabold\cdot\uv(\rv) - 2\partial_\mu\left[ \left( \partial_\alpha n_0(\rv) \right) u_{\nu\alpha}(\rv) \right]\right\},
\eer
($u_{\nu\mu} (\rv) \equiv(\partial_\nu u_\mu(\rv) +\partial_\mu u_\nu(\rv) )/2$ is the strain tensor)
and the interaction contribution is given by
\be\label{Fint}
{\cal F}^{\rm int}_{{\rm el},{\mu}}(\rv) = \int {\it d} \rv'  {\it K}_{\mu\nu}(\rv,\rv') [{\it u}_\nu(\rv)-{\it u}_\nu (\rv')]\,.
\ee
In Eq.~(\ref{Fk}), we have defined 
\ber\label{TMAGmunu0}
T^{\rm mag}_{\mu\nu,0}(\rv)&=&\frac{1}{2}\left(\partial_\mu\partial_\nu^\prime+\partial_\nu\partial_\mu^\prime\right) \gamma_0(\rv,\rv') \vert_{\rv=\rv'} - \frac{1}{4}\nabla^2 n_0(\rv)\delta_{\mu\nu}
\nn \\ &+& 3~n_0(\rv) v_{p0,\mu}(\rv) v_{p0,\nu}(\rv)  \,,
\eer
where $\gamma_0(\rv,\rv')$ is the ground-state one-particle density matrix [see Eq.~(\ref{gamma0})].
In Eq.~(\ref{Fint})
\be\label{CoulombKernel}
K_{\mu\nu}(\rv,\rv')=\Gamma_0(\rv,\rv')\partial_\mu\partial_\nu^\prime W(|\rv-\rv'|)\,,
\ee
where $W(|\rv-\rv'|)$ is the interaction potential and $\Gamma_0(\rv,\rv')$ is the ground-state pair distribution function [see Eq.~(\ref{Gamma0})].  

Analysis of the one-particle case within the Lagrangian formulation is reported in the Appendix C, thus this also completes the analysis started in the Appendix B within the Euler approach.

\section{Conclusions}\label{VI}
In summary, we recapitulate the essential changes that must be made in order to go from the quantum continuum mechanics in the absence of magnetic field to the one in the presence of magnetic field, within the framework of the ``elastic approximation": 
\begin{enumerate}

\item[(i)] A Lorentz-force term must be added to the equation of motion for the displacement field, and appropriately linearized, taking into account the presence of a non-vanishing velocity field $\vv_0(\rv)$ in the ground-state.

\item[(ii)] Everywhere, the time derivative $\partial_t$ must be replaced by the convective derivative   $D_t = \partial_t + \left( \vv_0 \cdot \nabla\right)$.  The replacement must also be done within the Lorentz force term, compatibly with the requirements of linearization. 

\item[(iii)]The kinetic contribution to the elastic energy must be calculated taking into account the replacement of the canonical momentum operator $-i\nabla$ by the
kinematic momentum $\hat{\bm\pi} = -i\nabla -\vv_{0 p}$. 

\item[(iv)]  The relation between linearized current density and displacement field is changed from Eq.~(\ref{Displacement1}) to Eq.~(\ref{Displacement2}).

\end{enumerate}

The Fourier transform of Eq.~(\ref{eom.1}) yields the following generalized eigenvalue problem
\ber\label{eomDisp3}
\omega^2 ~  \uv &+& i \omega ~  \left[   2  \left( \vv_0 \cdot \nabla \right) \uv +  \left( \uv \times \Bv_0 \right) \right] 
- \left( \vv_0 \cdot \nabla \right)^2 \uv  \nn \\ 
&-&  \left[   \left( \vv_0 \cdot \nabla  \uv \right) \times \Bv_0 \right]  - \left( \uv\cdot\nabla \right) \partial_\mu {\it V}_{\rm 0}(\rv) \nn \\ &-& \vv_0 \times  \left( \uv \cdot  \nabla \right) \Bv_{0}(\rv) 
+  \frac{1}{n_0} \fbold_{\rm el}= 0\;.
\eer
Finally,  we have  an explicit form of all the terms: this is a major step forward. 

In conclusion, we have presented  results that open the possibility to obtain the response of the current, and thus the excitation energies, of systems in strong magnetic fields
avoiding the solution of the time-dependent Schr\"odinger  equation and making use only of ground-state properties.  Given the required ground-state properties, 
the complexity of the problem  to be solved does not increase with the number of the particles in the system. The 
presented framework is expected to be useful in dealing with large systems and with current-carrying states exhibiting an elastic behavior.

\begin{acknowledgments}
S.P. and G.V were supported by DOE grant DE-FG02-05ER46203.
I.V.T. was supported by the Spanish MICINN, Grant No. FIS2010-21282-C02-01, and ``Grupos Consolidados UPV/EHU del Gobierno Vasco'', Project No. IT-319-07.
S.P. and G.V. acknowledge interesting discussions with Zeng-hui Yang about non-analytic time behavior of quantum states.
\end{acknowledgments}

 \appendix
 
 \section{Generalized eigenvalue problem} 
 
Eq.~(\ref{eomDisp-B}) can be represented in the form 
\be\label{NSEPD2}
\omega^2 ~ \tilde{\bf u}  +  i \omega ~ \widetilde{\calBv} \cdot \tilde{\bf u} - \widetilde{\calKv} \cdot \tilde{\bf u} =  0\;,
\ee
where we introduced the following notations
\ber\label{rename1}
\tilde{u}_\mu(\rv) &=& \sqrt{n_0(\rv)} u_{\mu}(\rv)\;, \\ \nn \\ \label{rename2}
\tilde{{\cal B}}_{\mu,\nu}(\rv,\rv') &=& i \frac{1}{\sqrt{n_0(\rv)}} {\cal B}_{\mu,\nu}(\rv,\rv')  \frac{1}{\sqrt{n_0(\rv')}}\;, \\ \nn \\ \label{rename3}
\tilde{{\cal K}}_{\mu \nu} (\rv,\rv') &=& \frac{1}{\sqrt{n_0(\rv)}} {\cal K}_{\mu,\nu}(\rv,\rv')  \frac{1}{\sqrt{n_0(\rv')}}\;.
\eer
Moreover, Eq.~(\ref{eomDisp3}) shares the same structure although (as we have explained) it is for a different dispalcement. 

The operators acting on $\widetilde{\uv}$ have properties
\be
\tilde{ {\cal B}}_{\mu,\nu}(\rv,\rv') = \tilde{ {\cal B}}^*_{\mu,\nu}(\rv,\rv'),~ \tilde{ {\cal B}}_{\mu,\nu}(\rv,\rv')  = - \tilde{ {\cal B}}_{\nu,\mu}(\rv',\rv)
\ee
and
\be
\tilde{ {\cal K}}_{\mu,\nu}(\rv,\rv')  = \tilde{ {\cal K}}^*_{\mu,\nu}(\rv,\rv') ,~ \tilde{ {\cal K}}_{\mu,\nu}(\rv,\rv')  =  \tilde{ {\cal K}}_{\nu,\mu}(\rv',\rv)\;.
\ee
Eq. (\ref{NSEPD2})  has the form of a non-standard eigenvalue problem which may be found for example also in the analysis of the 
modes of rotating stars \cite{Schenketal} and in the analysis of the magneto hydrodynamic of hot plasma \cite{FriemanRotenberg,Barton0}.  

In terms of the scalar product [the difference with Eq.~(\ref{product1}) is related to the rescaling of the displacement field, Eq.~(\ref{rename1})] 
\be
\langle  \tilde{\uv}_B,  \tilde{\uv}_A \rangle \equiv \int  d \rv ~  \tilde{u}^*_{B,\mu} (\rv)   \tilde{u}_{A,\mu}(\rv) \;,
\ee
the orthogonality relation get modified as follows 
\be\label{newo3}
\langle  \tilde{\uv}_B, i   \widetilde{\calBv} \cdot \tilde{\uv}_A \rangle + \left( \omega_A + \omega_B \right) \langle  \tilde{\uv}_B, \tilde{\uv}_A \rangle = 0\;,
\ee
where $\omega_A$ and $\omega_B$  are assumed to be real and different from each.
In fact, the operator defining the problem in Eq.~(\ref{NSEPD2}) is Hermitian for real $\omega$ but it is also $\omega$-dependent;
thus, solutions of  Eq. (\ref{rename1}) corresponding to different $\omega$ do not need to be orthogonal in the usual sense (as  for $\widetilde{\calBv} \equiv 0$).

Let us discuss the conditions for which the solutions are guaranteed to have real-valued $\omega$. 
For this, Eq.~(\ref{NSEPD2}) can be brought  the quadratic form
\be\label{stabilityE}
\omega^2 \langle \tilde{\uv}, \tilde{\uv} \rangle +  i~\omega \langle \tilde{\uv} ,  \widetilde{\calBv} \cdot \tilde{\uv} \rangle - \langle   \tilde{\uv}, \widetilde{\calKv} \cdot \tilde{\uv} \rangle = 0\;.
\ee
Real-valued $\omega$ are obtained for  positive definitive discriminant of Eq.~(\ref{stabilityE}).
Since $\langle  \tilde{\uv},  \tilde{\calBv} \cdot \tilde{\uv} \rangle$ is a purely imaginary quantity, the stability condition may be stated in terms of the stronger requirement
\be\label{stabilityC}
\langle \tilde{\uv}, \widetilde{\calKv } \cdot \tilde{\uv} \rangle > 0\;.
\ee
Eq.~(\ref{stabilityC}) is not manifestly satisfied for the equations under considerations.
Nevertheless, for
Eq.~(\ref{eomDisp1}) and  Eq.~(\ref{eomDisp3}), we expected to find stable solutions because 
those same equations are valid for small displacements and short-time intervals: 
within these conditions, the system must stay ``close" to the the initial minimum (the ground-state) of the unperturbed Hamiltonian.

\section{One-particle case in the Euler description with the standard displacement} \label{oneparticle1}
Here, we show that the inversion of the current response for  one particle system can be easily worked out from the linearized Schroedinger equation or, equivalently, from the
linearized Euler equations for the densities. In this way, we are able to show that the approximation put forward in the high-frequency limit 
provide the {\em exact} excitation energies for one particle systems.

We start by observing that, the wave function may be written as follows
\be\label{1WF}
\Psi = \sqrt{n} e^{i \varphi}\;,
\ee
thus,
\be\label{1C}
\jv = n \nabla \varphi + n\Av\;,
\ee
and
\be\label{1VR}
\nabla \times \vv =  \Bv\;.
\ee
As a result,
the local balance equation for the linear momentum reads as follows
\be\label{1EEQ}
\partial_t j_\mu +  \partial_\mu [ \;  {\cal V}_B + \frac{1}{2} \frac{j^2}{n^2} + V \; ] =  \partial_t A_\mu
\ee
where
\be\label{VB}
 {\cal V}_B = - \frac{1}{2} \frac{\nabla^2 \sqrt{n}}{\sqrt{n}}
\ee
is the well-known Bohm potential~\cite{Bohm,Takabayashi}. 
Let 
\be
V = V_0\;,~~~\mbox{and}~~~ \Av = \Av_0 + \Av_{1}\;,
\ee
the linearization of  Eq.~(\ref{1EEQ}) yields
\ber\label{L1EEQ}
\frac{ \partial_t \jv_{1}}{n_0} &+& \frac{  \left( \nabla \cdot \jv_1 \right) }{n_0}  \vv_{0}
+ \nabla \left\{ {\cal V}_{1,B} + \vv_0 \cdot \frac{\jv_1}{n_0}  - v_0^2 \frac{ n_1 }{n_0} \right\} \nn \\
&=& \partial_t \Av_{1}
\eer
where
\be
{\cal V}_{1,B}  = - \frac{1}{4}  \left[  \frac{\nabla^2 \frac{n_1}{\sqrt{n_0}}}{\sqrt{n_0}}  -  \frac{ \nabla^2 \sqrt{n_0}   }{\sqrt{n_0}}  \left(   \frac{n_1}{n_0}  \right) \right]\;,
\ee
moreover, we may also remind that
\be\label{1CE}
\partial_t n_1 = - \nabla \cdot \jv_{1}\;.
\ee

From Eq.~(\ref{L1EEQ}), we obtain an important relation for the inverse of the current-current response function:
\ber\label{INV}
{\bm \chi}^{-1}(\omega) \cdot \jv_1 &=& 
\frac{\jv_1}{n_0} + \frac{i}{\omega} \left[ \left( \nabla \cdot \jv_1 \right)  \frac{\vv_0}{n_0} +  \nabla \left( \vv_0 \cdot \frac{\jv_1}{n_0} \right) \right] \nn \\
&+& \frac{1}{\omega^2}  \frac{\nabla}{2} \left[ \frac{1}{\sqrt{n_0}} \left( - \frac{\nabla^2}{2} + \frac{ \nabla^2 \sqrt{n_0} }{2\sqrt{n_0}} \right. \right. \nn \\ 
&-&  \left. \left. 2 v^2_0 \right) \frac{\nabla \cdot \jv_1}{\sqrt{n_0}} \right] \;.
\eer
This expression tells us that the frequency dependency includes only $1/\omega$ and $1/\omega^2$ terms: therefore, it is apparent that
the inversion in high-frequency limit  of the current response function is {\em exact} for one particle systems.

Let us consider  purely longitudinal perturbation, from Eq.~(\ref{L1EEQ})  we get the equation for modes described in terms of $\jv_1 = - i \omega n_0 \uv$ 
\ber\label{1PEP}
\omega^2 {\uv} &+& 
i \omega \left[ \left( \nabla \cdot  n_0 {\uv} \right)  \frac{\vv_0}{n_0} + \nabla \left( \vv_0 \cdot {\uv} \right) \right] \nn \\
&+& \frac{\nabla}{2} \left[ 
\frac{1}{\sqrt{n_0}} \left(- \frac{\nabla^2}{2} + \frac{ \nabla^2 \sqrt{n_0} }{2\sqrt{n_0}}   - 2 v^2_0 \right) \frac{\nabla \cdot \left( n_0 {\uv} \right) }{\sqrt{n_0}}  \right] \nn \\
&=& \nabla V_1 \;,
\eer
Now, we find the {\em exact} displacement for the one-particle system. For this purpose, we use the expression of the linearized Schr\"odinger equation.
Without loosing generality, one may shift to zero the energy of ground-state energy and choose the ground-state wave function in Eq.~(\ref{1WF}) to be a real-valued function (i.e.; $\varphi_0 =  0$). 
In the given gauge, the relation 
\be
\Av_0 = \frac{\jv_0}{n_0} = \vv_0 \;
\ee
holds true. By substitution in Eq.~(\ref{1PEP}), we can verify that the solutions have form
\be
{\uv} = \frac{\left[\; \jv \; \right]_{0n}}{n_0},~~~ \omega=\omega_{n0}
\ee
where
\be
\left[\;\jv\;\right]_{0n} = - \frac{i}{2} \Psi^2_0 \; \nabla \left( \frac{\Psi_n}{\Psi_0} \right) + \Psi_0 \Psi_n \Av_0\;.
\ee
are the matrix element of the current operator evaluated for the eigenstates, $\Psi_n$, of $\hat{H}_0$ and  $\omega_{n0}$ 
are the corresponding excitation energies.
In verifying the above results, it may be useful to remind  the continuity relations
\be\label{cont1}
\nabla \cdot \left[\;\jv\;\right]_{0n} = i \omega_{n0} \Psi_0 \Psi_n\;.
\ee

\section{One-particle case in the Lagrangian description with the new displacement} \label{oneparticle1L}

Let us consider the one-particle case but within the elastic approximation as obtained within the Lagrangian description.
In this case, we
readily arrive at
\begin{eqnarray} \label{Ekin-N=1}
E_{{\rm el}}^{N=1}[g_{\mu\nu}] &=& \int d\xiv ~\frac{1}{2}\sqrt{g}g^{\mu\nu}  \nn \\ 
&\times&  \left(\partial_{\mu} \sqrt{\frac{n_0}{g^{1/2}}} \right)\left(\partial_{\nu}\sqrt{\frac{n_0}{g^{1/2}}}\right)\;.
\end{eqnarray}
It is remarkable that,  Eq.~(\ref{Ekin-N=1}) has the same form as in the limit of vanishing magnetic field.
Moreover,  it is possible to directly verify that the Lagrangian with the elastic energy as in Eq.~(\ref{Ekin-N=1}):
\begin{eqnarray}
\label{Lagr-magn-N=1}
L &=& \int d\bm\xi n_0(\bm\xi)\left[\frac{1}{2}(D_t\xv)^2 - \left( D_t\xv \right) \Av_0(\xv) - V(\xv,t) \right] \nn \\
&-& \frac{\sqrt{g}g^{\mu\nu}}{2}\left(\partial_{\mu}\sqrt{\frac{n_0}{g^{1/2}}} \right)\left(\partial_{\nu}\sqrt{\frac{n_0}{g^{1/2}}}\right) 
\end{eqnarray}
yields the {\it exact} equation of motion for the trajectory $\rv(\xiv,t)$. In other words, 
the elastic approximation is {\em exact} for one-particle systems regardless the value of $\Bv_0$.

Then, the  linearized equation of motion is readily obtained
\ber\label{leom1}
\ddot{u}_\mu &+& v_{0, \nu} \partial_\nu \dot{u}_\mu - \dot{u}_\nu \partial_\nu v_{0, \mu} + \partial_\mu \left( v_{0,\nu} \dot{u}_\nu \right) \nn \\
&+& \partial_\mu \left\{ 
v_{0,\alpha}v_{0,\beta} u_{\alpha \beta} +  \frac{\nabla^2 u_{\alpha \beta}}{4} +  \frac{\partial_\alpha u_{\alpha \beta} \partial_\beta \sqrt{n_0}}{\sqrt{n_0}}  \right. \nn \\ 
&+& \left.
u_\alpha \partial_\alpha  V_0 
\right\} = \partial A_{1,\mu}\;.
\eer
The latter expression can also be obtained from Eq.~(\ref{1EEQ})  by re-expressing the response of the current in terms of 
``Lagrangian" displacement
\be
j_{1,\mu} = n_0 \dot{u}_\mu + \partial_\nu \left( j_{0,\nu} u_{\mu} - j_{0,\mu} u_{\nu} \right)\;.
\ee
In conclusion,  as expected, the same equation is obtained if the same displacement is employed consistently.


\begin{thebibliography}{99}
\bibitem{footnote}
A typical example is a situation when the ground-state density has cusps, as for example in the  important case in which the static external potential is the Coulomb potential of the nucleus $Ze/r$, producing a cusp in the density at $r=0$.  In this case the response to external perturbation, for example, the total inelastic cross section, exhibit non-analytic behavior at high frequency~\cite{ZMK11}.
A number of other examples of non-analytic short-time behavior generated by spatially non-analytic initial states can be found in Refs.~\onlinecite{HS71,MTWB10}.
\end{thebibliography}
\end{document}